\documentclass[preprint,aps,amsmath,nofootinbib,floatfix]{revtex4}

\usepackage{supertabular}
\usepackage{supertabular}
\usepackage{graphicx}
\usepackage{epstopdf}
\usepackage{subfigure}
\usepackage{multirow}

\usepackage{slashed}

\usepackage{color}
\usepackage{comment}
\usepackage{booktabs} 
\usepackage{natbib}
\usepackage[colorlinks,linkcolor=black,anchorcolor=blue,citecolor=green]{hyperref}
\makeatletter

\newcommand{\Rmnum}[1]{\expandafter\@slowromancap\romannumeral #1@}
\makeatother
\allowdisplaybreaks

\begin{document}
\title{Four-body decays of $B$ meson with lepton number violation}
\author{
        Han Yuan$^{1}$\footnote{hanyuan@hit.edu.cn},
        Tianhong Wang$^{1}$\footnote{thwang@hit.edu.cn},
        Yue Jiang$^1$\footnote{jiangure@hit.edu.cn},
        Qiang Li$^1$\footnote{lrhit@protonmail.com}
        Guo-Li Wang$^{1}$ \footnote{gl\_wang@hit.edu.cn},
         }

\affiliation{
$^{1}$ Department of Physics, Harbin Institute of Technology, Harbin, 150001, China
}

\begin{abstract}
This paper is designed to calculate the branching ratio of four-body decays of $B$ meson with lepton number changed by 2. With the new experimental data limit to lepton-number violation processes, we update the upper limits of mixing parameters between heavy Majorana neutrino and charged leptons. Afterwards, we calculate the branching ratio of   $B^0(P)\rightarrow D^{*-}(P_1)\ell^+_1(P_2)\ell^+_2(P_3)M_2^-(P_4)$ using the updated parameters. It is found that the most hopeful decay channel is $B^0(P)\rightarrow D^{*-}(P_1)e^+_1(P_2)e^+_2(P_3)\rho^-(P_4)$ or $B^0(P)\rightarrow D^{*-}(P_1)e(\mu)^+_1(P_2)\mu(e)^+_2(P_3)\rho^-(P_4)$, whose branching ratio can reach about $10^{-4}$ with heavy Majorana neutrino mass range around $2~\mathrm{GeV}$.
\end{abstract}
\maketitle

\section{Introduction}
As a nature of Majorana field, Majorana neutrino and its antineutrino is the same. The corresponding Lagrangian also shows that processes which violate lepton number are possible. So the search for lepton-number violation (LNV) process is known as the best way to determine whether the neutrino is Dirac or Majorana fermion \cite{Pas2015}. 

The most attractive searches for LNV are the processes of the neutrinoless double-beta decays of nucleus \cite{GomezCadenas:2011it}. Processes involving mesons or $\tau$ lepton decaying to LNV final states have been suggested as alternative ways. Some experiments have already searched for such LNV processes for many years \cite{Agostini2013,Collaboration2014,Gando2013}. BaBar Collaboration \cite{Lees2011,Lees2012BarBar,Lees2014} show their experimental results about processes $M\rightarrow M _1\ell_1\ell_2$, where $M$ can be $D, ~D_s,~ \Lambda^+_c,~B^+$, $\ell$ is $e$ or $\mu$, $M_1$ can be meson $\pi, ~K,~ \rho, ~K^*, ~D$, etc. LHCb \cite{Aaij2012,Aaij20122,Aaij2014} present results of $|\Delta L|=2$ decays of $B$ and $D_{(s)}$. Belle Collaboration \cite{Seon2011,Miyazaki2013} separately display the $0\nu\beta\beta$ channels of $B^+$ and $\tau^-$. Future experiments also have the potential to find more possible results, Belle II \cite{Wang2015,Aushev2010} purpose to get $\sim 40$ times more data than Belle I, and LHC Run 2 will also be updated. All these experiments will help us to search for the hidden $|\Delta L|=2$ processes. On the other way, the meson $|\Delta L|=2$ processes can also supply the information of mixing parameters between the Majorana neutrino and charged leptons. So the observation of meson $|\Delta L|=2$ processes would show both the existence of Majorana neutrino and the data of Majorana neutrino parameters.

Theoretically, in high energy range, LNV $|\Delta L|=2$ decays of mesons and $\tau$ have also been studied carefully in literature \cite{Helo2011,Cvetic2010,Zhang2011,Bao2013,Wang2014,Milanes2016,Mandal2016,Delepine2011,Castro2012,LopezCastro2013,Dong2013,Yuan2013,Castro20122,Dib2012,Yuan:2017xdp,Kim:2017pra}. The meson decay processes can be roughly separated into two kinds, one is three-body channels $M\rightarrow M _1\ell_1\ell_2$ or  $\tau\rightarrow\ell M_1M_2$\cite{Helo2011,Cvetic2010,Zhang2011,Bao2013,Wang2014,Milanes2016}; the other one is four-body channels $M\rightarrow M _1\ell_1\ell_2M_2$ or $\tau\rightarrow\ell M_1M_2\nu$\cite{Milanes2016,Mandal2016,Delepine2011,Castro2012,LopezCastro2013,Dong2013,Yuan2013,Castro20122,Dib2012}; where $M$ can be $K, D, D_s, B$ or $B_c$, $\ell$ represents $e,~\mu,~\tau$,  $\nu$ is the corresponding neutrino of $\ell$, $M_1$ and $M_2$ are $\pi$, $K$,  $\rho$, $K^*$, $D$, or $D_s$, which are the final mesons allowed by kinematics.

\begin{figure}[!hbt]
\centering
\subfigure[]{\includegraphics[scale=0.45]{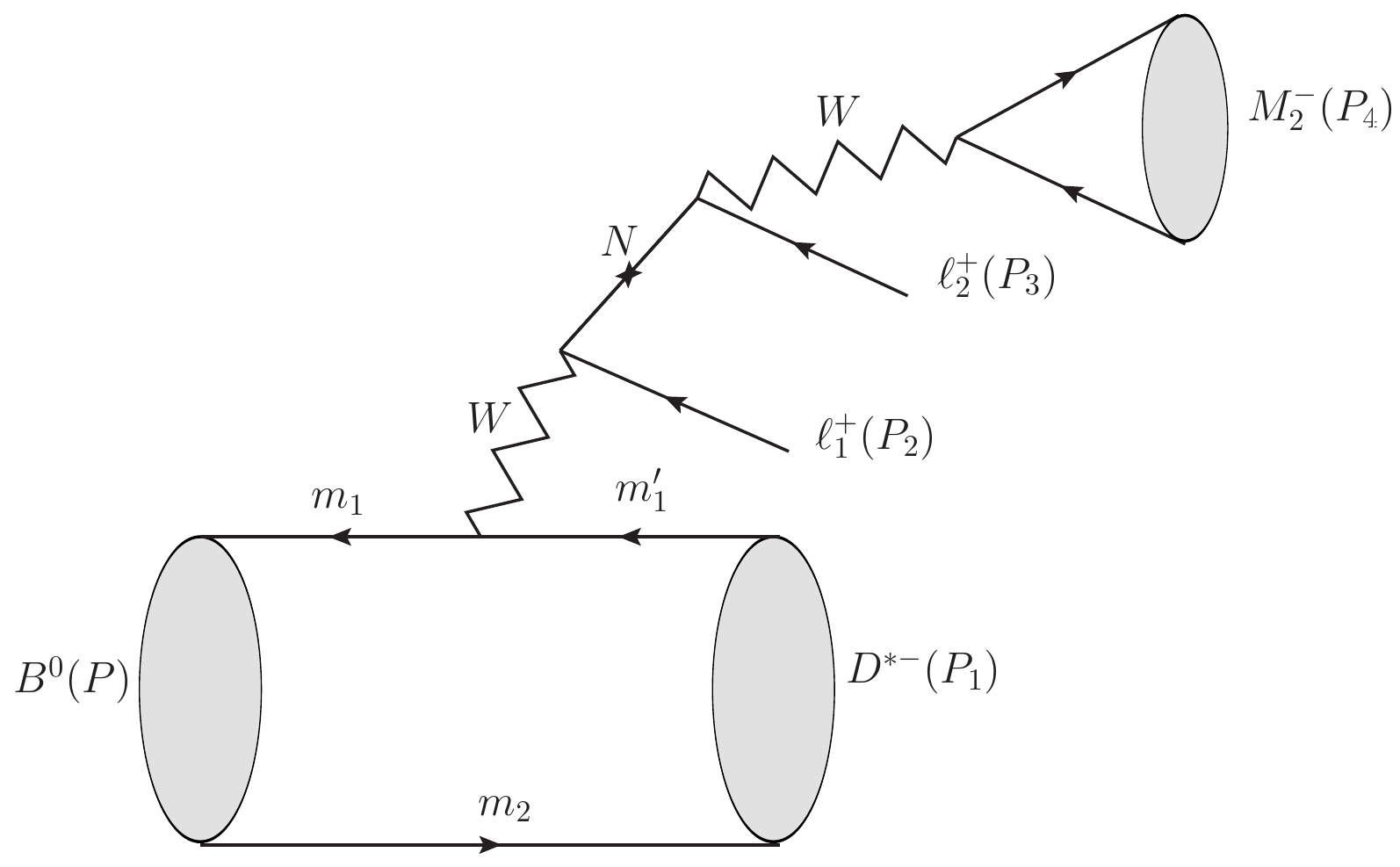}\label{fig:Feyn1}}
\subfigure[]{\includegraphics[scale=0.45]{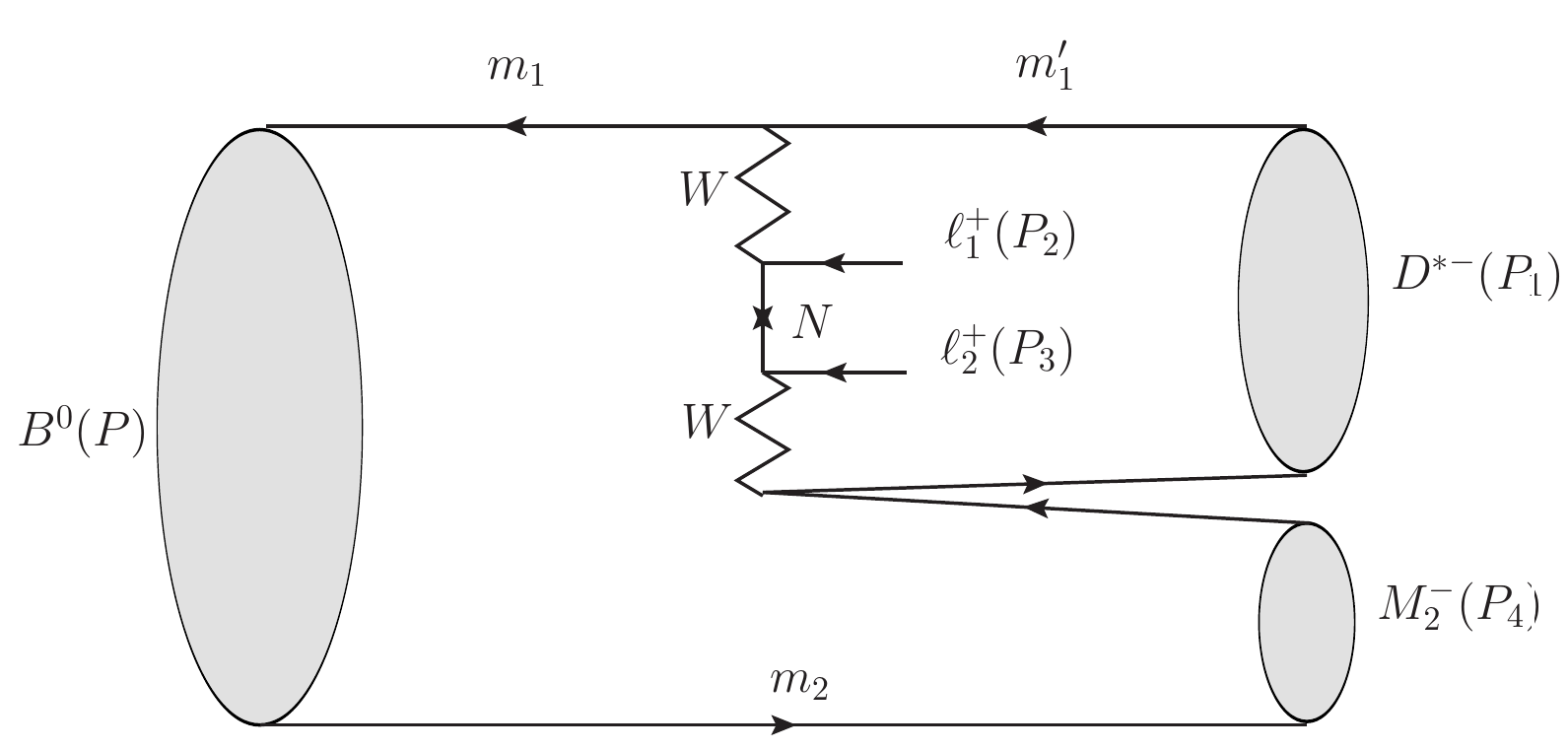}\label{fig:Feyn2}}
\caption{Feynman Diagram of $B^0(P)\rightarrow D^{*-}(P_1)\ell^+_1(P_2)\ell^+_2(P_3)M_2^-(P_4)$.}\label{fig:Feyn}
\end{figure}

In our previous paper \cite{Yuan2013}, heavy pseudoscalar mesons $B$ and $D$ four-body $|\Delta L|=2$  decays have been studied. For $B$ meson processes $B\rightarrow M _1\ell_1\ell_2M_2$, the final meson $M_1$ is a pseudoscalar meson. Since the four-body LNV or lepton-flavor violation (LFV) decays of $B$ share the same vertexes and mixing parameters as well as the CKM matrix elements with the corresponding three-body $B$ decays, thus the branching ratios of four-body LNV (LFV) processes are not much smaller than the corresponding three-body one, and have sizable results, so as a kind of extension, we add vector $M^{*}_1$ in this paper. The channel we focus on is $B^0(P)\rightarrow D^{*-}(P_1)\ell^+_1(P_2)\ell^+_2(P_3)M_2^-(P_4)$ (where $M_2$ is another meson), whose Feynman diagram is shown in Fig.\ref{fig:Feyn}.
We can see that there are two diagrams of this process. In Fig.\ref{fig:Feyn1}, if the neutrino mass is between several MeV and GeV, the internal neutrino can be on mass-shell, which is a heavy neutrino. Because of neutrino-resonance, on-shell neutrino will greatly enhance the branching ratio of the decays, which make such kind of decays are detectable in current experiments. In Fig.\ref{fig:Feyn2} the neutrino is off-shell, whose branching ratio will be much smaller than the previous one, thus we only consider the process of Fig.\ref{fig:Feyn1}.

Some parameters should be considered in calculation, such as the Majorana neutrino mass, the Majorana neutrino decay width, the mixing parameters between neutrino and charged lepton. Since we let the heavy Majorana neutrino be on mass-shell, so the mass range is determined by kinematics. For the mixing parameters between Majorana neutrino and charged leptons, we use the latest experimental data to get the upper limits of them. The decay width of heavy Majorana neutrino with a certain mass can be got in the same way as Ref. \cite{Atre2009}.

This work is organized as follows: In Sec. II we present the procedure of our calculation which includes the decay amplitude and some useful parameters. In Sec. III we give the results and discussions. Our conclusions are listed in Sec. IV. Some details about the  hadronic matrix elements are put in the Appendix A.
\section{Theoretical and calculation detail}
The Feynman diagram of $B^0(P)\rightarrow D^{*-}(P_1)\ell^+_1(P_2)\ell^+_2(P_3)M_2^-(P_4)$ is shown in Fig.\ref{fig:Feyn1}, where meson $B$ with momentum $P$, $D^{*-}$ with momentum $P_1$, $\ell_1$($\ell_2$) is lepton $e$ or $\mu$  with momentum $P_2$($P_3$), charged meson $M_2$ can be pseudoscalar meson $\pi$, $K$, $D$ and $D_s$ or vector meson $\rho$ and $K^*$, whose momentum is $P_4$. According to the Feynman rules in Ref. \cite{Atre2009}, which is a simple extension of Standard Model, we can get the transition matrix element of $W^+W^+\rightarrow\ell^+\ell^+$
\begin{eqnarray}
\mathcal{L}_{\mu\nu}&=&\frac{g^2}{2}\sum\limits_{m=1}\limits^{3}U_{\ell_1m}^{l\nu}U_{\ell_2m}^{l\nu}m_{\nu_m}\\\nonumber
&&\times\bar{u}_1\left(\frac{\gamma_\mu\gamma_\nu}{q^2-m_{\nu m}^2+i\Gamma_{\nu m}m_{\nu m}}+\frac{\gamma_\nu\gamma_\mu}{q^{\prime2}-m_{\nu m}^2+i\Gamma_{\nu m}m_{\nu m}}\right)P_R\nu_2\\\nonumber
&&+\frac{g^2}{2}\sum\limits_{m^\prime=4}\limits^{3+n}V_{\ell_1m^\prime}^{lN}V_{\ell_2m^\prime}^{lN}m_{N_{m^\prime}}\\\nonumber
&&\times\bar{u}_1\left(\frac{\gamma_\mu\gamma_\nu}{q^2-m_{N_{m^\prime}}^2+i\Gamma_{N_{m^\prime}}m_{N_{m^\prime}}}+\frac{\gamma_\nu\gamma_\mu}{q^{\prime2}-m_{N_{m^\prime}}^2+i\Gamma_{N_{m^\prime}}m_{N_{m^\prime}}}\right)P_L\nu_2,\\\nonumber
\end{eqnarray}
where $U$ and $V$ (dropped the superscripts) are mixing parameter matrixes between neutrino and charged leptons, $m_{\nu m}$ ($m_{N_m^\prime}$) is neutrino mass, $q$ ($q^\prime$) is momentum of neutrino, $\Gamma$ is decay width of neutrino, $u$ ($v$) is spinor of lepton, $P_R(L)=\frac{1}{2}(1 \pm \gamma_5)$. For the light Majorana neutrinos, $m = 1,~2,~3$, whose masses are very light. And for heavy Majorana neutrino, $m^\prime = 4\sim~3+n$. If the heavy Majorana mass satisfies $m_{N_{m^\prime}}\approx q^2~(q^{\prime2})$, the heavy Majorana neutrino process will has a resonant enhancement. However the light Majorana neutrino contribution will be suppressed by small neutrino mass. So we drop the $\sum_{m=1}^3$ part of transition matrix element. All the heavy Majorana neutrinos have contributions to the amplitude $\mathcal{L}_{\mu\nu}$, but in our analysis we only consider the contribution of one of the heavy neutrinos in particular the lightest one for simplicity $N=4$. In the process $B^0(P)\rightarrow D^{*-}(P_1)\ell^+_1(P_2)\ell^+_2(P_3)M_2^-(P_4)$, the heavy Majorana neutrino is on mass-shell, whose mass is in the range of $\mathrm{MeV}\sim\mathrm{GeV}$ allowed by kinematics, and this process can be separated into two parts $B^0(P)\rightarrow D^{*-}(P_1)\ell^+_1(P_2)N$ and $N\rightarrow \ell^+_2(P_3)M_2^-(P_4)$. The two leptons can be distinguished as they are in different processes, so the exchange of two leptons in $B^0(P)\rightarrow D^{*-}(P_1)\ell^+_1(P_2)\ell^+_2(P_3)M_2^-(P_4)$ do not need to be considered. The amplitude $\mathcal{L}_{\mu\nu}$ can be written as
\begin{equation}\label{eq:Delta}
\mathcal{L}_{\mu\nu}=\frac{g^2}{4}V_{\ell_14}V_{\ell_24}m_4\bar{u}(P_2)\frac{\gamma_\mu\gamma_\nu}{q^2-m_4^2+i\Gamma_4m_4}(1-\gamma_5)v(P_3).
\end{equation}
Similarly, the two meson $M_1$ and $M_2$ are in different sub-processes, so we also do not need consider the exchange of these two mesons.

The transition amplitude for the 4-body decay $B^0(P)\rightarrow D^{*-}(P_1)\ell^+_1(P_2)\ell^+_2(P_3)M_2^-(P_4)$ shown in Fig.\ref{fig:Feyn1} can be written as:
\begin{equation}\label{eq:amp}
\mathcal{M}=\frac{g^2V_{q_1q_2}V_{q_3q_4}}{8M_W^2}\langle D^{*-}(P_1)\vert\bar{q}_1\gamma^\mu(1-\gamma_5)q_2\vert B^0(P)\rangle\times
\mathcal{L}_{\mu\nu}\times\langle M_2(P_4)\vert\bar{q}_3\gamma^\nu(1-\gamma_5)q_4\vert0\rangle
\end{equation}
where the momentum dependence in the propagator of $W$ boson has been ignored since it is much smaller than the $W$ mass; $V_{q_1q_2}(V_{q_3q_4})$ is the Cabibbo-Kobayashi-Maskawa (CKM) matrix element between quarks $q_1$ and $q_2$ ($q_3$ and $q_4$); $\mathcal{L}_{\mu\nu}$ is the transition amplitude of the leptonic part.

The corresponding hadronic matrix element can be described as a function of form factors \cite{Wang:2009as}:
\begin{eqnarray}\label{eq:hadronic}
&&\langle D^{*-}(P_1)\vert\bar{q}_2\gamma_\mu(1-\gamma_5)q_1\vert B^0(P)\rangle\\\nonumber
&&=f\epsilon^\lambda_{1\mu}+(a_++a_-)(P\cdot\epsilon_1^\lambda)P_\mu+(a_+-a_-)(P\cdot\epsilon_1^\lambda)P_{1\mu}+ig^\prime\varepsilon_{\mu\nu\rho\sigma}\epsilon_1^{\lambda\nu}(P+P_1)^\rho(P-P_1)^\sigma,
\end{eqnarray}
where $\epsilon_1$ is polarization vector of $D^{*-}$, $f$, $a_+$, $a_-$ and $g^\prime$ are form factors. The details about how to calculate them are listed in appendix.

The last part $\langle M_2(P_4)\vert\bar{q}_3\gamma^\nu(1-\gamma_5)q_4\vert0\rangle$ in Eq.\eqref{eq:amp} is related to the decay constant of the meson $M_2$. If $M_2$ is a pseudoscalar with momentum $P_4$, we have the following relation:
\begin{equation}\label{eq:constant}
\langle M_2(P_4)|{\bar{q}_3}\gamma^\nu(1-\gamma_5)q_4|0\rangle=i F_{M_2}P_4^\nu,
\end{equation}
where $F_{M_2}$ is decay constant of meson $M_2$. If $M_2$ is a vector with momentum $P_4$ and polarization vector $\epsilon$, then the corresponding relation becomes
\begin{equation}\label{eq:constant1}
\langle M_2(P_4,\epsilon)|{\bar{q}_3}\gamma^\nu(1-\gamma_5)q_4|0\rangle=M_2F_{M_2}{\epsilon}^\nu,
\end{equation}
where we use the same symbol $M_2$ to denote the meson and its mass.

Combining Eq.\eqref{eq:Delta}, Eq.\eqref{eq:hadronic} and Eq.\eqref{eq:constant},  we can get the final amplitude of $B^0(P)\rightarrow D^{*-}(P_1)\ell^+_1(P_2)\ell^+_2(P_3)M_2^-(P_4)$ in the case of meson $M_2$ being as a pseudoscalar
\begin{eqnarray}
i\mathcal{M}&=&\frac{iG_F^2V_{q_1q_2}V_{q_3q_4}V_{\ell_14}V_{\ell_24}m_4F_{M2}}{q^2-m_4^2+i\Gamma_4m_4}\\\nonumber
&&\times\bar{u}(P_2)\left[f\slashed{\epsilon}^\lambda_{1\mu}+(a_++a_-)(P\cdot\epsilon_1^\lambda)\slashed{P}_\mu+(a_+-a_-)(P\cdot\epsilon_1^\lambda)\slashed{P}_{1\mu}\right.\\\nonumber
&&\left.+ig^\prime\varepsilon_{\mu\nu\rho\sigma}\epsilon_1^{\lambda\nu}(P+P_1)^\rho(P-P_1)^\sigma\gamma_\mu\right]\slashed{P_4}(1-\gamma_5)v(P_3).
\end{eqnarray}
The amplitude of $B^0(P)\rightarrow D^{*-}(P_1)\ell^+_1(P_2)\ell^+_2(P_3)M_2^-(P_4)$ in the case of meson $M_2$ being as a vector can be written in the same way.

\begin{figure}[!hbt]
\subfigure[Excluded region above the curves for $|V_{e4}|^2$ versus $m_4$ from $M_1^+\rightarrow e^+e^+M_2^-$.]{\includegraphics[scale=0.26]{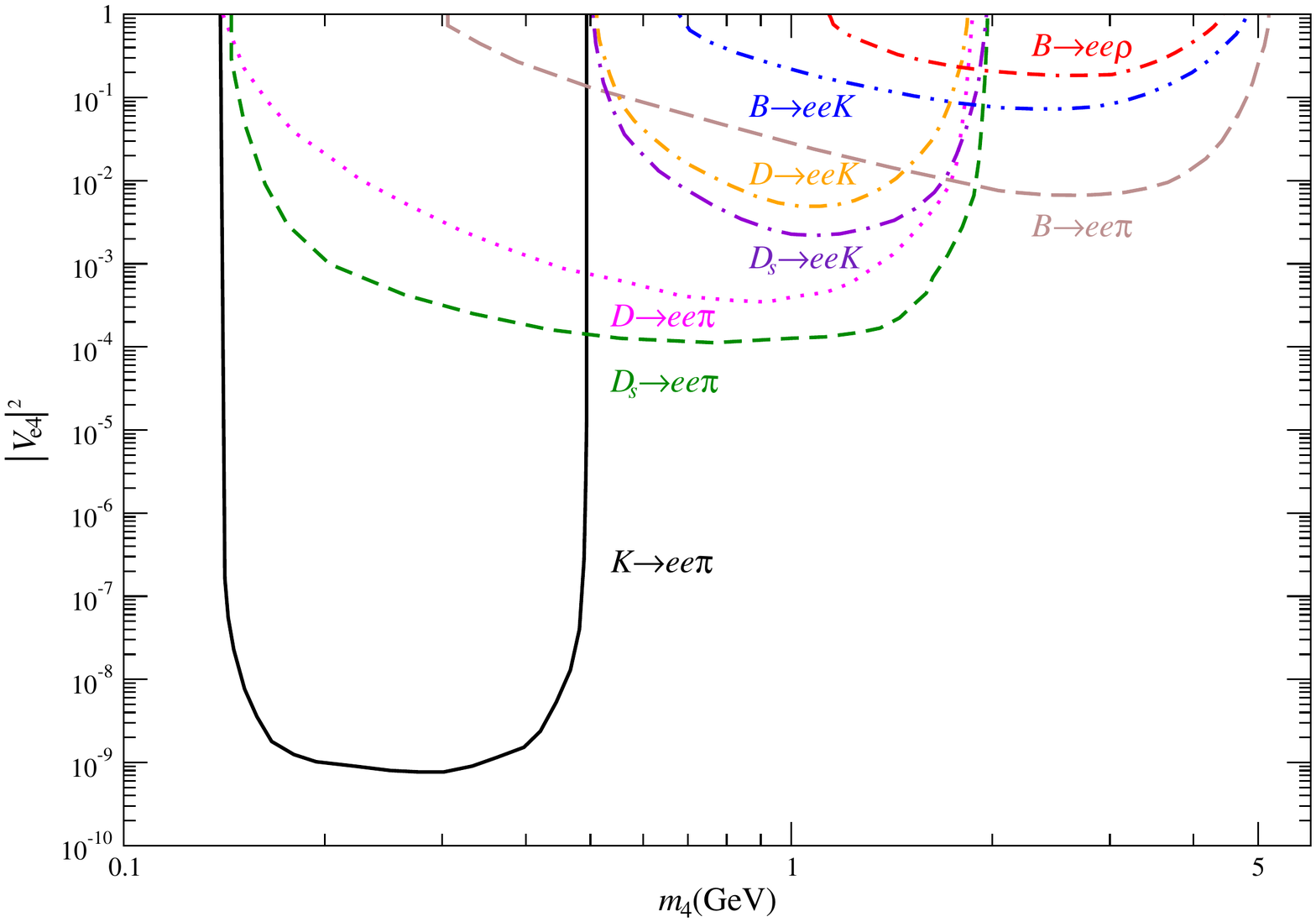}\label{fig:mpee}}
\subfigure[Excluded region above the curves for $|V_{e4}V_{\mu4}|$ versus $m_4$ from $M_1^+\rightarrow e^+\mu^+M_2^-$.]{\includegraphics[scale=0.26]{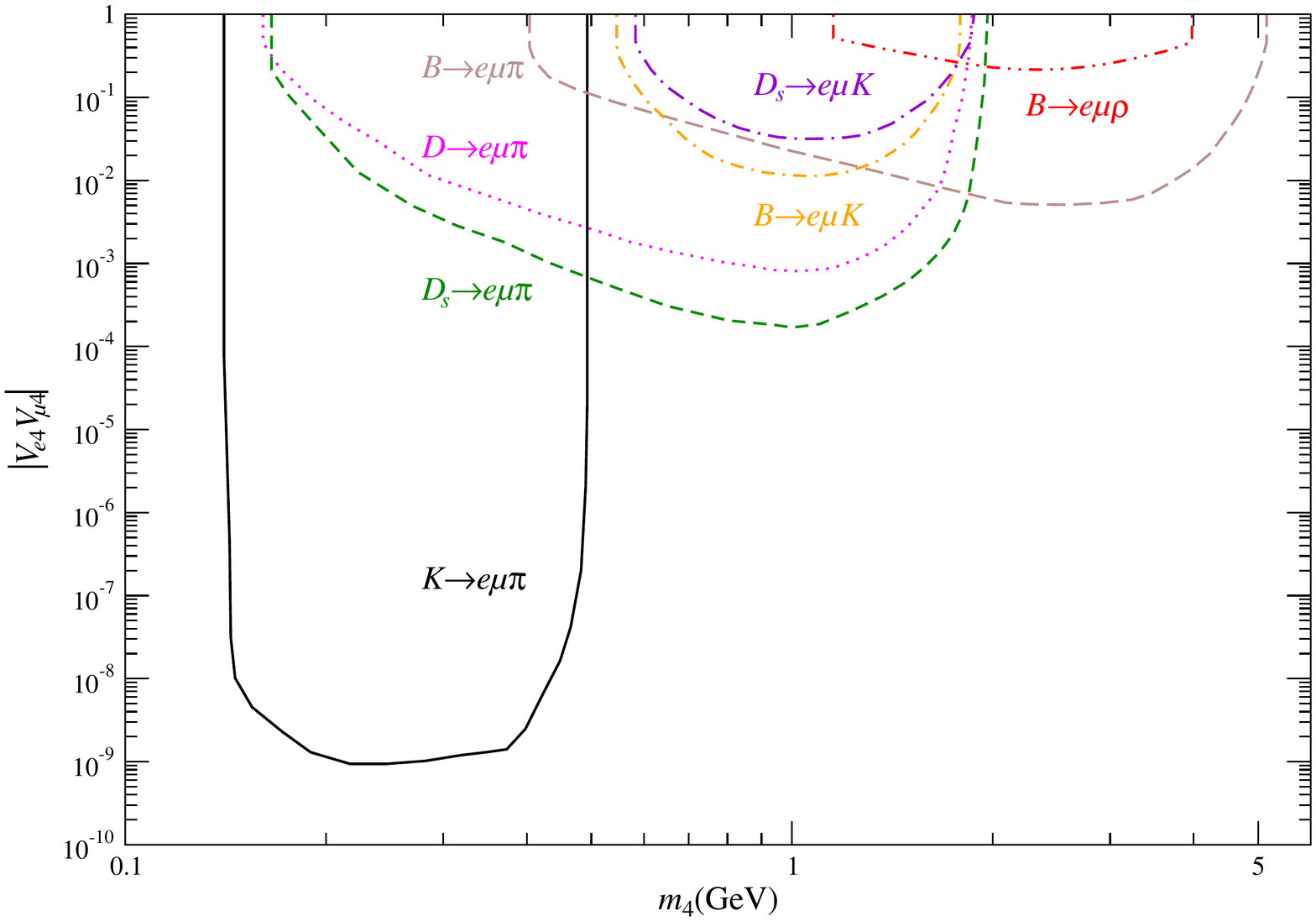}\label{fig:mpemu}}
\subfigure[Excluded region above the curves for $|V_{\mu4}|^2$ versus $m_4$ from $M_1^+\rightarrow \mu^+\mu^+M_2^-$.]{\includegraphics[scale=0.26]{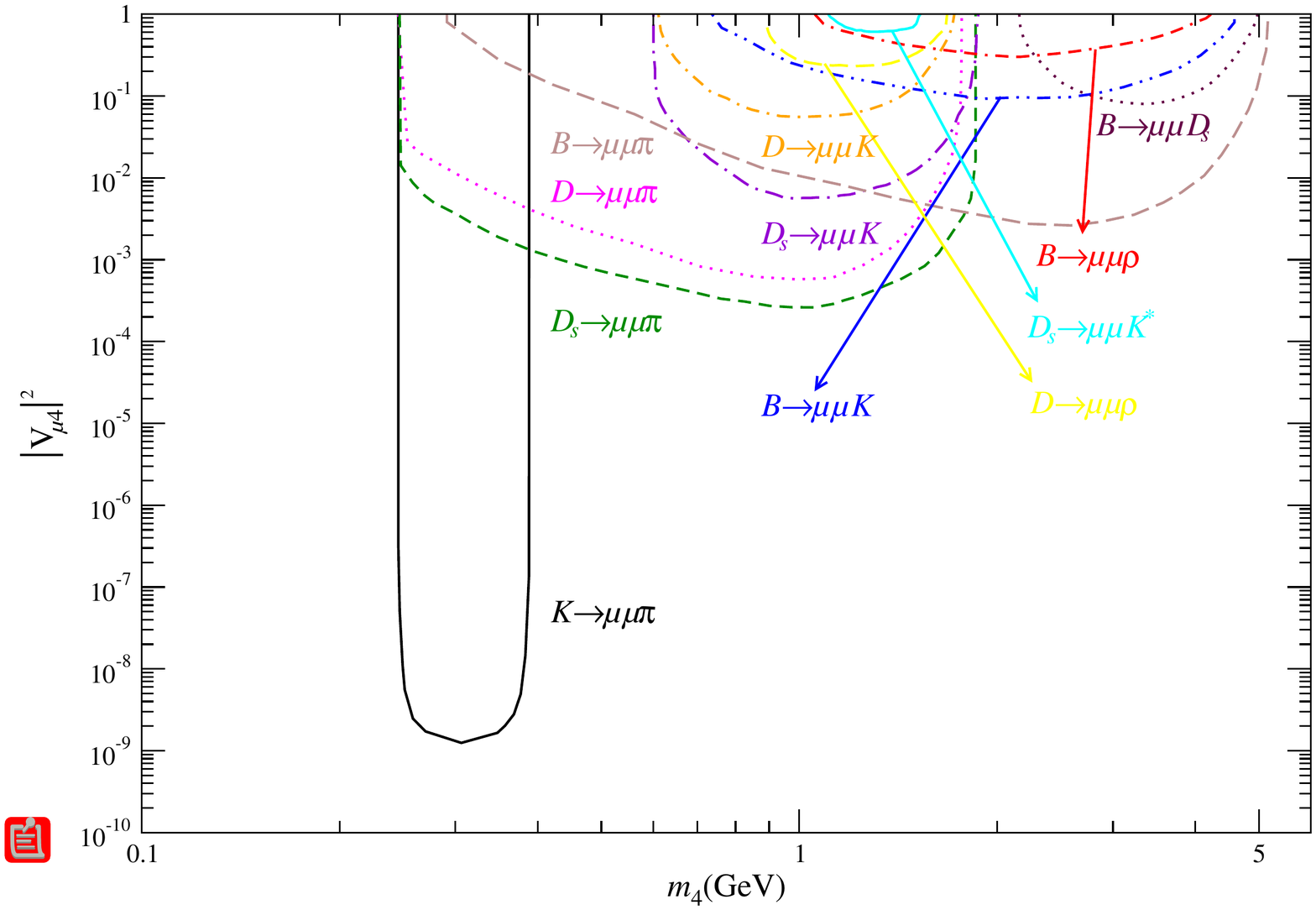}\label{fig:mpmumu}}
\caption{The region in the curves are excluded by experimental data of LNV (LFV) channels with the range of $m_4$ coming from $M_1\rightarrow\ell\ell M_2$ searches.}\label{fig:mp}
\end{figure}

There are some important input parameters in calculation, such as $V_{\ell4}$, $m_4$ and $\Gamma_4$. For the $V_{\ell4}$ we follow the method of \cite{Atre2009} and take advantage of the experimental data from \cite{Agashe:2014kda,Aaij2012} to give new limits for $|V_{\ell_14}V_{\ell_24}|$, which is shown in Fig.\ref{fig:mp}.
The range of $m_4$ is determined by kinematics and we followed the method of Ref. \cite{Atre2009} to calculate all the possible decay channels of heavy Majorana neutrino to get its decay width.

With all setting parameters the decay width of $B^0(P)\rightarrow D^{*-}(P_1)\ell^+_1(P_2)\ell^+_2(P_3)M_2^-(P_4)$ can be calculated as follows
\begin{eqnarray}
\Gamma&=&\frac{1}{2M}\int\frac{\mathrm{d}^3P_1}{(2\pi)^32E_1}\frac{\mathrm{d}^3P_2}{(2\pi)^32E_2}\frac{\mathrm{d}^3P_3}{(2\pi)^32E_3}\frac{\mathrm{d}^3P_4}{(2\pi)^32E_4}\\\nonumber
&&\times(2\pi)^4\delta^4(P-P_1-P_2-P_3-P_4)|\mathcal{M}|^2\\\nonumber
&=&\frac{1}{512\pi^8M}\int\mathrm{d}_4(ps,P\rightarrow P_1P_2P_3P_4)|\mathcal{M}|^2.
\end{eqnarray}
Since the Majorana neutrino is on mass-shell, the 4-body phase space integral can be converted to the product of 2-body phase space integrals
\begin{equation}
\mathrm{d}_4(ps,P\rightarrow P_1P_2P_3P_4)=\mathrm{d}_2(ps,P\rightarrow P_1P_W)\mathrm{d}M_W^2\mathrm{d}_2(ps,P_W\rightarrow P_2x)\mathrm{d}M_x^2\mathrm{d}_2(ps,x\rightarrow P_3P_4),
\end{equation}
where $P_W$ is the momentum of $W$, $x$ is the momentum of heavy Majorana neutrino, $M_x$ is mass of heavy Majorana neutrino. The universal 2-body phase space integral can be treated as follows,
\begin{equation}
\mathrm{d}_2(ps,X\rightarrow ab)=\frac{\pi}{2}\lambda^{\frac{1}{2}}(1,\frac{a^2}{X^2},\frac{b^2}{X^2})\frac{\mathrm{d}\Omega}{4\pi},
\end{equation}
where $X$ is the momentum of initial state particle, $a$ and $b$ are the momentums of final particles, $\lambda$ function is
\begin{equation}
\lambda(x,y,z)=x^2+y^2+z^2-2xy-2yz-2xz.
\end{equation}
Because the decay width of heavy Majorana neutrino is very small, we use Narrow Width Approximation
\begin{equation}
\left.\int\frac{dM_x^2}{(M_x^2-m_4^2)^2+\Gamma_{4}^2m_4^2}\right|_{\Gamma_{4}^2\rightarrow0}=\int\delta\left(M_x^2-m_4^2\right)\mathrm{d}M_x^2\frac{\pi}{\Gamma_{4}m_4}.
\end{equation}
Although the 4-body phase space integral has been separated to the product of 2-body phase space integral, it is still very complex, so we use Monte Carlo method to get the value of this integral. 
\section{Results and analysis}
We choose the CKM matrix elements \cite{Agashe:2014kda} as follows: $V_{ud}=0.974$, $V_{us}=0.225$, $V_{cd}=0.225$, $V_{cs}=0.986$, $V_{cb}=41.1\times10^{-3}$, $V_{ub}=4.13\times10^{-3}$.
The decay constants of pseudoscalar and vector mesons used in our calculation are listed in Table~\ref{tab:const}.
\begin{table}[h]
\caption{ Decay constants $F_{M_2}$ of pseudoscalar and vector mesons in unit of MeV.}
\begin{center}
{\begin{tabular}{|c|c|c|c|c|c|c|c|} \hline
meson&$\pi$&$\rho$&$K$&$K^{*}$&$D$&$D_s$&$D^{*}$\\\hline
$F_{M_2}$&130.4~\cite{Agashe:2014kda}&220~\cite{Ebert:2006hj}&156.1~\cite{Agashe:2014kda}&217~\cite{Ebert:2006hj}&222.6~\cite{Artuso:2005ym}&260~\cite{Agashe:2014kda}&310~\cite{Ebert:2006hj}\\\hline
\end{tabular} \label{tab:const}}
\end{center}
\end{table}
We use the experimental upper limit of branching fraction of LNV (LFV) meson decay to get the relationship between mixing parameters $|V_{\ell_14}V_{\ell_24}|$ and heavy Majorana neutrino $m_4$. And the decay channels used are listed in Table \ref{tab:mp}.
\begin{center}
\tablefirsthead{%
	\hline
	\multicolumn{1}{|c}{Mixing parameter} &
	\multicolumn{1}{|c}{Mass Range (MeV)} &
	\multicolumn{1}{|c}{Decay mode} &
	\multicolumn{1}{|c|}{\emph{Branching Ratio}}\\
	\hline}
\tablehead{%
	\hline
	\multicolumn{4}{|r|}{continued from previous page}\\
	\hline
	\multicolumn{1}{|c}{Mixing parameter} &
	\multicolumn{1}{|c}{Mass Range (MeV)} &
	\multicolumn{1}{|c}{Decay mode} &
	\multicolumn{1}{|c|}{\emph{Branching Ratio}}\\
	\hline}
\tabletail{%
	\hline
	\multicolumn{4}{|r|}{continued on next page}\\
	\hline}
\tablelasttail{\hline}
\topcaption{Mixing parameters and mass range to corresponding LNV (LFV) meson decay \cite{Agashe:2014kda,Aaij20122}.}
\begin{supertabular}{|c|c|c|c|}
&140-493&$K^+\rightarrow e^+e^+\pi^-$&$6.4\times10^{-10}$\\\cline{2-4}
&140-1868&$D^+\rightarrow e^+e^+\pi^-$&$1.1\times10^{-6}$\\\cline{2-4}
&494-1868&$D^+\rightarrow e^+e^+K^-$&$9.0\times10^{-7}$\\\cline{2-4}
&140-1967&$D_s^+\rightarrow e^+e^+\pi^-$&$4.1\times10^{-6}$\\\cline{2-4}
$|V_{e4}|^2$
&494-1967&$D_s^+\rightarrow e^+e^+K^-$&$5.2\times10^{-6}$\\\cline{2-4}
&140-5278&$B^+\rightarrow e^+e^+\pi^-$&$2.3\times10^{-8}$\\\cline{2-4}
&494-5278&$B^+\rightarrow e^+e^+K^-$&$3.0\times10^{-8}$\\\cline{2-4}
&776-5278&$B^+\rightarrow e^+e^+\rho^-$&$2.6\times10^{-6}$\\\cline{2-4}
&892-5278&$B^+\rightarrow e^+e^+K^{*-}$&$2.8\times10^{-8}$\\\hline
&140-493&$K^+\rightarrow e^+\mu^+\pi^-$&$5.0\times10^{-10}$\\\cline{2-4}
&140-1868&$D^+\rightarrow e^+\mu^+\pi^-$&$2.0\times10^{-6}$\\\cline{2-4}
&494-1868&$D^+\rightarrow e^+\mu^+K^-$&$1.9\times10^{-6}$\\\cline{2-4}
&140-1967&$D_s^+\rightarrow e^+\mu^+\pi^-$&$8.4\times10^{-6}$\\\cline{2-4}
$|V_{e4}V_{\mu4}|$
&494-1967&$D_s^+\rightarrow e^+\mu^+K^-$&$6.1\times10^{-6}$\\\cline{2-4}
&140-5278&$B^+\rightarrow e^+\mu^+\pi^-$&$1.3\times10^{-6}$\\\cline{2-4}
&494-5278&$B^+\rightarrow e^+\mu^+K^-$&$2.0\times10^{-6}$\\\cline{2-4}
&776-5278&$B^+\rightarrow e^+\mu^+\rho^-$&$3.3\times10^{-6}$\\\cline{2-4}
&892-5278&$B^+\rightarrow e^+\mu^+K^{*-}$&$4.4\times10^{-6}$\\\hline
&245-388&$K^+\rightarrow \mu^+\mu^+\pi^-$&$1.1\times10^{-9}$\\\cline{2-4}
&245-1763&$D^+\rightarrow \mu^+\mu^+\pi^-$&$2.0\times10^{-6}$\\\cline{2-4}
&599-1763&$D^+\rightarrow \mu^+\mu^+K^-$&$1.0\times10^{-5}$\\\cline{2-4}
&881-1763&$D^+\rightarrow \mu^+\mu^+\rho^-$&$5.6\times10^{-4}$\\\cline{2-4}
&997-1763&$D^+\rightarrow \mu^+\mu^+K^{*-}$&$8.5\times10^{-4}$\\\cline{2-4}
$|V_{\mu4}|^2$
&245-1862&$D_s^+\rightarrow \mu^+\mu^+\pi^-$&$1.4\times10^{-5}$\\\cline{2-4}
&599-1862&$D_s^+\rightarrow \mu^+\mu^+K^-$&$1.3\times10^{-5}$\\\cline{2-4}
&997-1862&$D_s^+\rightarrow \mu^+\mu^+K^{*-}$&$1.4\times10^{-3}$\\\cline{2-4}
&245-5173&$B^+\rightarrow \mu^+\mu^+\pi^-$&$1.3\times10^{-8}$\\\cline{2-4}
&599-5173&$B^+\rightarrow \mu^+\mu^+K^-$&$4.1\times10^{-8}$\\\cline{2-4}
&881-5173&$B^+\rightarrow \mu^+\mu^+\rho^-$&$5.0\times10^{-6}$\\\cline{2-4}
&2074-5173&$B^+\rightarrow \mu^+\mu^+D_s^-$&$5.8\times10^{-6}$\\\hline
\end{supertabular}\label{tab:mp}
\end{center}

The last column in Table \ref{tab:mp} is the upper limit of branching ratio of LNV (LFV) decay in experiments \cite{Agashe:2014kda,Aaij20122}. The branching fraction of some decay modes like $K^+\rightarrow e^+e^+\pi^-$ keeps constant when compared with \cite{Atre2009}, as a result, some curves in Fig.\ref{fig:mpee}, Fig.\ref{fig:mpemu} and Fig.\ref{fig:mpmumu} are the same with those in \cite{Atre2009}. However, many processes have changed a lot, and the limit of mixing parameters $|V_{\ell_14}V_{\ell_24}|$ would change correspondingly. In Fig.\ref{fig:mpee} the position of curve $D^+\rightarrow e^+e^+\pi^-$ swaps with $D_s^+\rightarrow e^+e^+\pi^-$, which appears since the decay width of $D_s^+\rightarrow e^+e^+\pi^-$ changes a lot. Meanwhile the limit about $|V_{e4}|^2$ in the $m_4$ range from $0.5$ GeV to $2$ GeV changes from $10^{-3}$ to $10^{-4}$.  Another obvious transformation is the curve of $B^+\rightarrow e^+e^+\pi^-$, whose above region is bigger than the previous one, it means stricter limit about $|V_{e4}|^2$ corresponding to heavy Majorana neutrino mass range. Similarly evident variation happens in Fig.\ref{fig:mpemu}, the curves of $D^+\rightarrow e^+\mu^+\pi$ and $D^+_s\rightarrow e^+\mu^+\pi$ are separated, which also provide rigorous limit about $|V_{e_4}V_{\mu_4}|$ in $m_4$ ranged from $0.5$ GeV to $2$ GeV, the limit obtained from $D^+_s\rightarrow e^+\mu^+\pi$ decreases from $10^{-2}$ to $10^{-4}$, $B^+\rightarrow e^+\mu^+\pi^-$ also improves limit about $|V_{e_4}V_{\mu_4}|$ from $10^{-1}$ to $10^{-2}$ in $m_4$ range of $2\sim5$ GeV. In Fig.\ref{fig:mpmumu}, the most obvious variation appears in $B^+\rightarrow \mu^+\mu^+\pi^-$, which makes strict $|V_{\mu4}|^2$ in $m_4$ range from 2 to 5 GeV. To sum up, we provide stricter limits about mixing parameters $|V_{\ell_14}V_{\ell_24}|$ compared with the Ref.\citep{Atre2009}, which will help us to deal with more $|\Delta L|=2$ decays.

\begin{figure}[!hbt]
\subfigure[$B^0(P)\rightarrow D^{*-}(P_1)e^+_1(P_2)e^+_2(P_3)M_2^-(P_4)$]{\includegraphics[scale=0.4,angle = 0]{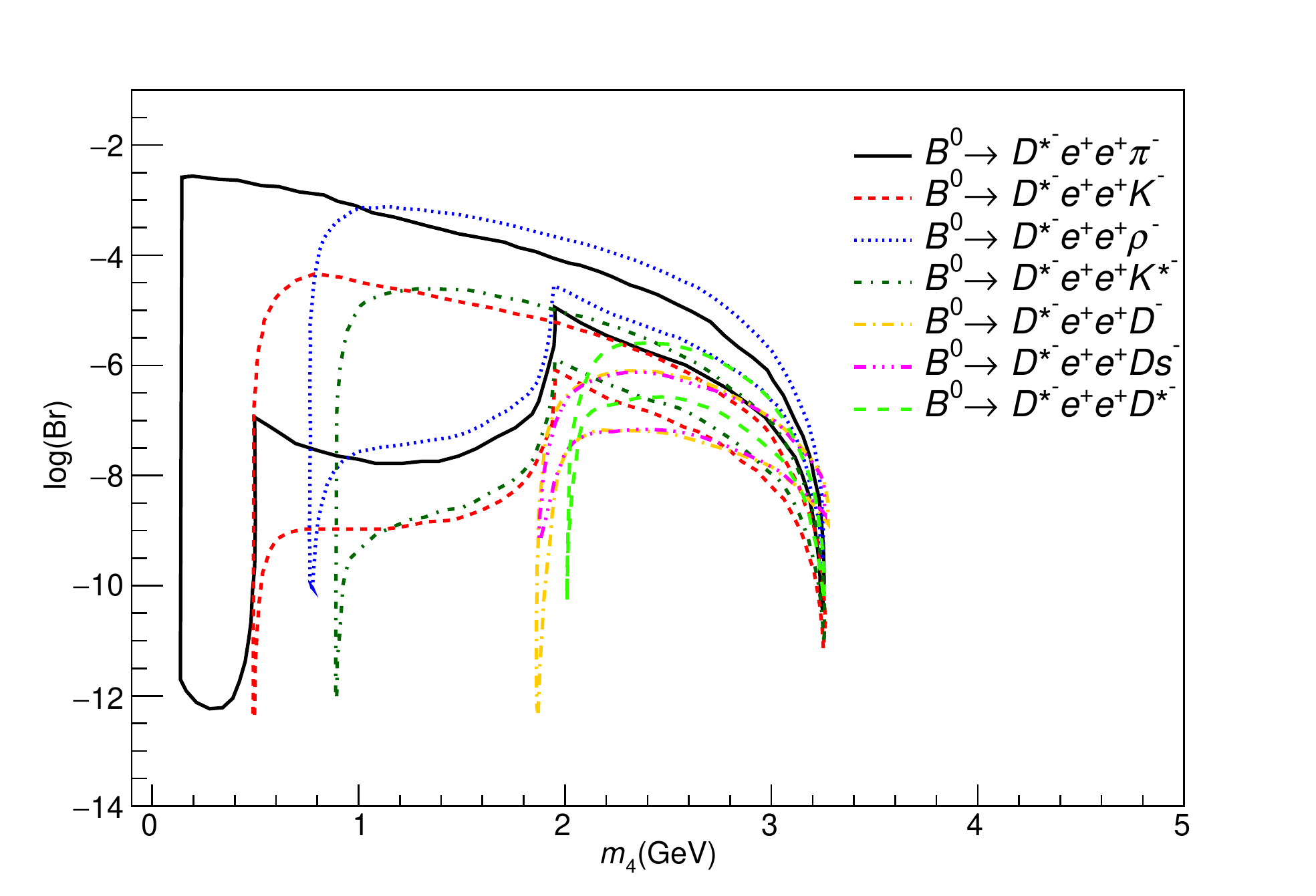}\label{fig:Bree}}
\subfigure[$B^0(P)\rightarrow D^{*-}(P_1)\mu^+_1(P_2)\mu^+_2(P_3)M_2^-(P_4)$]{\includegraphics[scale=0.4,angle = 0]{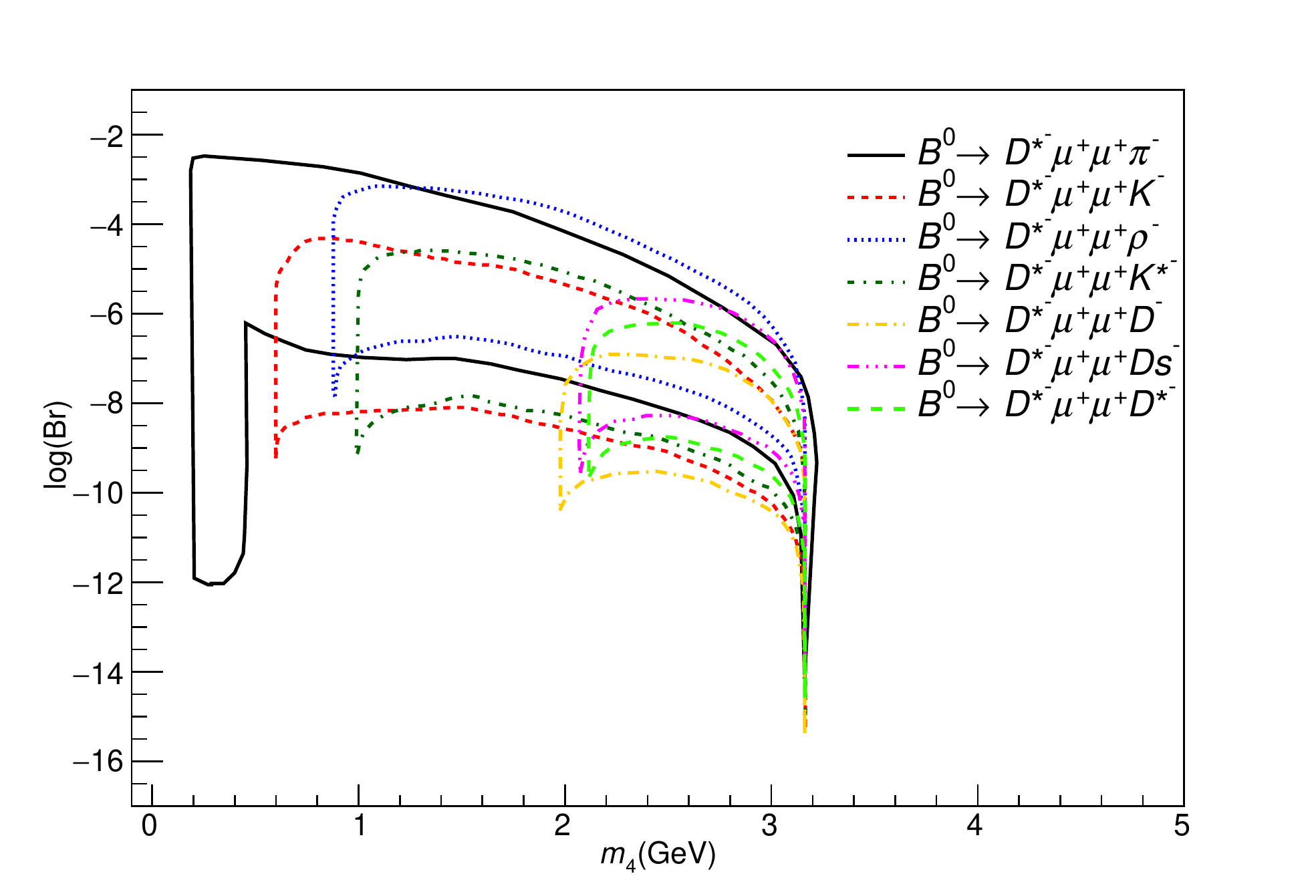}\label{fig:Brmm}}\\
\subfigure[$B^0(P)\rightarrow D^{*-}(P_1)e^+_1(P_2)\mu^+_2(P_3)M_2^-(P_4)$]{\includegraphics[scale=0.4,angle = 0]{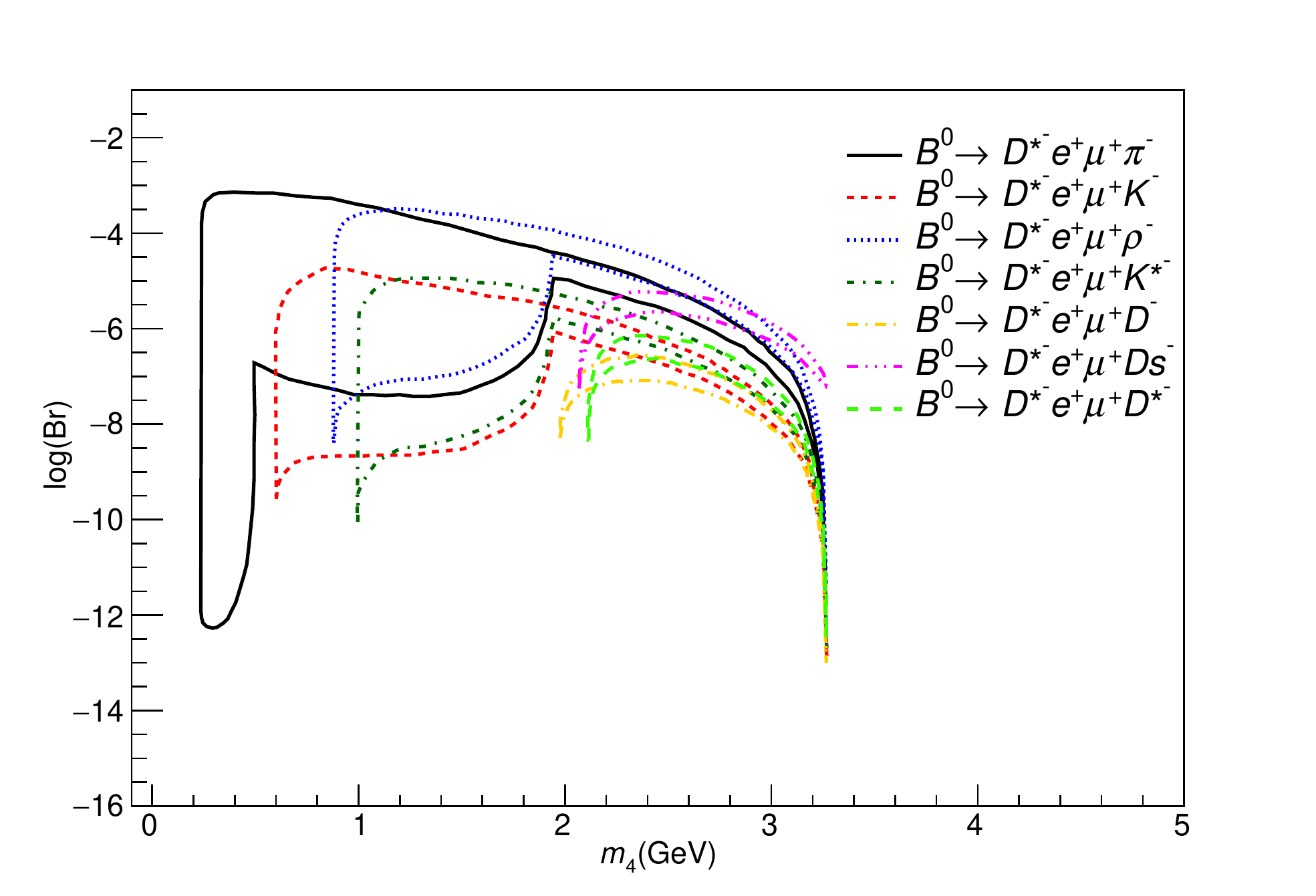}\label{fig:Brem}}
\subfigure[$B^0(P)\rightarrow D^{*-}(P_1)\mu^+_1(P_2)e^+_2(P_3)M_2^-(P_4)$]{\includegraphics[scale=0.4,angle = 0]{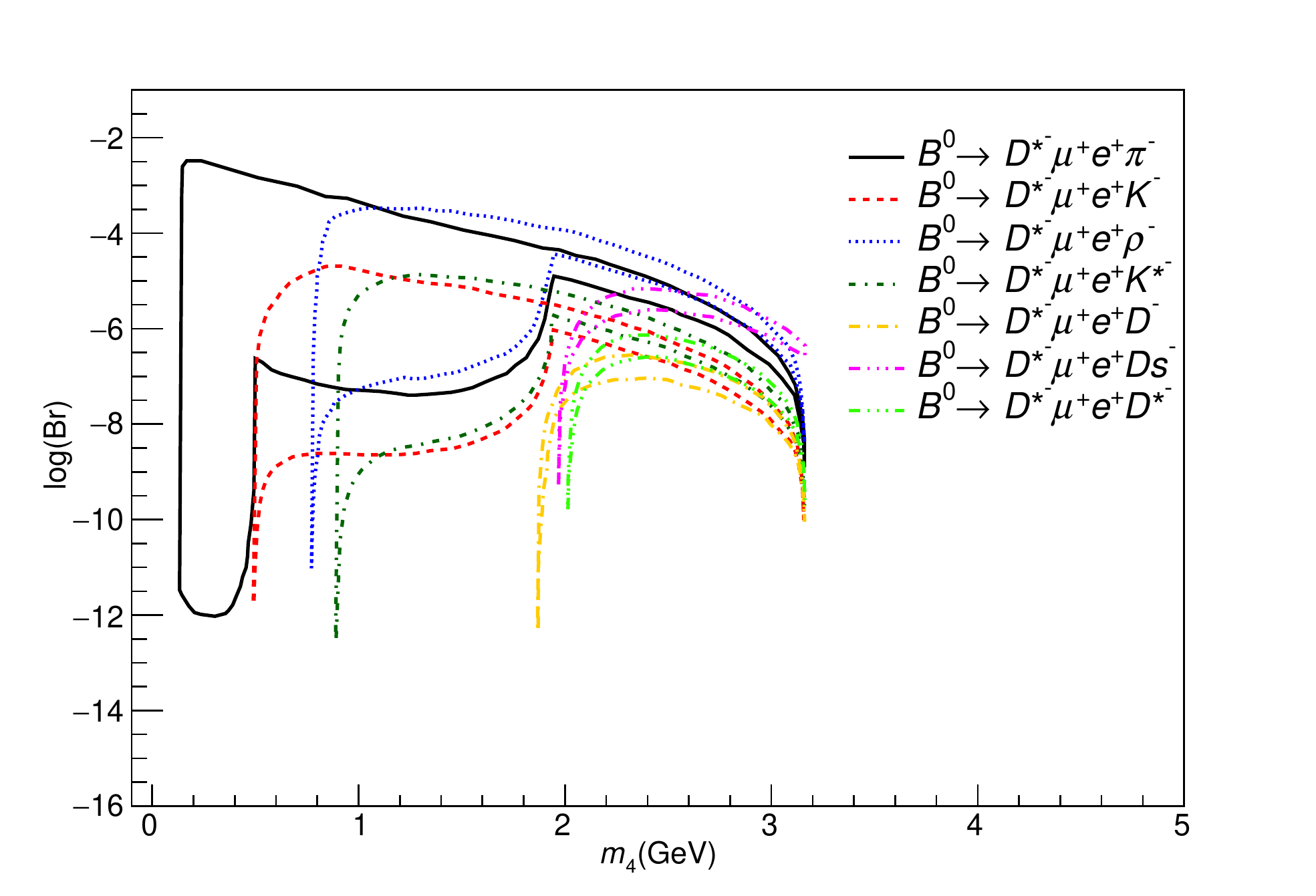}\label{fig:Brme}}
\caption{Function of heavy Majorana neutrino mass and $B^0(P)\rightarrow D^{*-}(P_1)\ell^+_1(P_1)\ell^+_2(P_3)M_2^-(P_4)$ decay $\log(Br)$, different color lines represent different decay channels. The region in the curves are exclude by theory.}\label{fig:Br}
\end{figure}

Then we focus on the branching ratio of $B^0(P)\rightarrow D^{*-}(P_1)\ell^+_1(P_2)\ell^+_2(P_3)M_2^-(P_4)$. In Fig.\ref{fig:Br},
the $x$-axis represents mass of heavy Majorana neutrino, and the $y$-axis is the $\log$ of branching ratio. We draw the relation between them on the canvas. In all three sub-figures in Fig.\ref{fig:Br}, the areas in and above the curves are excluded in theory as the points in these region with the mixing parameters $|V_{\ell4}|$ are not allowed in experiment. The sharp decreasing at the both two ends of the curve are because of the limits of phase space. In the following discussion all the analyses are about the lower edge of the curves. Because of limits of mixing parameters when heavy neutrino mass is smaller than $2~\mathrm{GeV}$ the $Br$ is small and hard to be observed. Also because of the piecewise mixing parameters the $Br$ curves are separated to several blocks. In Fig.\ref{fig:Br}, we choose the most stringent limits of mixing parameters in different heavy neutrino mass range. For example, when we calculate $B^0(P)\rightarrow D^{*-}(P_1)e^+_1(P_2)e^+_2(P_3)M_2^-(P_4)$ processes, if the heavy Majorana mass is smaller than $0.5~\mathrm{GeV}$, we use the limits of mixing parameters coming from $K\rightarrow ee\pi$. Similarly when heavy Majorana mass range is in $0.5~\mathrm{GeV}\sim2~\mathrm{GeV}$, the limits of mixing parameters are chosen from $D_s\rightarrow ee\pi$. The conspicuous difference of this two limits cause the curves of $Br$ are discontinuous. In Fig.\ref{fig:Brmm}, the curve of $B^0\rightarrow D^{*-}\ell^+\ell^+\pi^-$ is separated into two parts which is different from the processes in Fig.\ref{fig:Bree}, Fig.\ref{fig:Brem} and Fig.\ref{fig:Brme}, it is because the limits of mixing parameters are different in different pictures. If the two leptons in the process are both $\mu$, the limit comes from Fig.\ref{fig:mpmumu}. In Fig.\ref{fig:mpmumu}, the limit provided by decay channel $D_s\rightarrow\mu\mu\pi$ is about $10^{-3}$ in the heavy neutrino mass range $0.4$ GeV $\sim$ $2$ GeV. When heavy neutrino mass is lager than $2$ GeV, the limit mostly comes form $B\rightarrow\mu\mu\pi$ which is very close to the above one. But the situation is different in other mixing parameters pictures, like in Fig.\ref{fig:mpee}, the limit in $m_4$ range $0.5$ GeV $\sim$ $2$ GeV is about $10^{-4}$, and when $m_4~>2$ GeV the limit becomes $10^{-2}$, the difference is bigger than Fig.\ref{fig:mpmumu}, so the curve in Fig.\ref{fig:Bree} is separated into two parts when $m_4$ is larger than $0.5$ GeV, but in Fig.\ref{fig:Brmm} it is almost continuous. When the two leptons are different, the phase space is different, so the exchange leptons correspond different processes and their phase space are also different, so we draw them in different figures.

In Fig.\ref{fig:Br}, the area under the curves is allowed in theory. We can see in Fig.\ref{fig:Bree}, the largest branching ratio of $B^0(P)\rightarrow D^{*-}(P_1)e^+_1(P_1)e^+_2(P_3)\rho^-(P_4)$ can reach about $10^{-4}\sim10^{-6}$ with heavy neutrino mass between $2~\mathrm{GeV}\sim3~\mathrm{GeV}$, which is sizable in experiment. Fig.\ref{fig:Brmm} shows that in large heavy neutrino mass range $0.6$ GeV $\sim$ 3 GeV the largest branching ratio of $B^0\to D^{*-}\mu^+\mu^+\rho^-$ can reach about $10^{-7}\sim~10^{-8}$. Considering every year Belle Collaboration produced $772$ million $B\bar{B}$ events \cite{Seon2011} which can be used to study the four-body $B$ meson LNV and LFV decays. The $B^0$ decays with two final state $\mu$ leptons are hard to be detected in experiments. As the reason of mixing parameters between heavy Majorana neutrino and charged leptons, in Fig.\ref{fig:Brem} and Fig.\ref{fig:Brme} when heavy Majorana neutrino mass is larger than $2~\mathrm{GeV}$, the branching ratio can achieve about $10^{-5}$. As a conclusion, the processes in Fig.\ref{fig:Bree}, Fig.\ref{fig:Brem} and Fig.\ref{fig:Brme} with heavy Majorana neutrino mass around $2~\mathrm{GeV}$ have possibility to be found in the experiment, and with future update experiment the possibility will be enhanced.
\section{Summary}
In this paper, we update the upper limits of mixing parameters between charged leptons and heavy Majorana neutrino with experimental data,  we get the relationship between the Majorana neutrino mass and decay branching ratios and point out that some $|\Delta L|=2$ heavy meson 4-body decays have the possibility to be detected in current and future experiments.
\section*{Acknowledgments}
We would like to thank Tao Han for his suggestions to carry out this research and providing the FORTRAN codes Hanlib for the calculations. This work was supported in part by the National Natural Science Foundation of China (NSFC) under grant No.~11405037, 11575048 and 11505039.

\appendix
\renewcommand{\appendixname}{Appendix}
\section{Hadronic matrix element and form factors}
\begin{figure}[!hbt]
\centering
\includegraphics[scale=0.5]{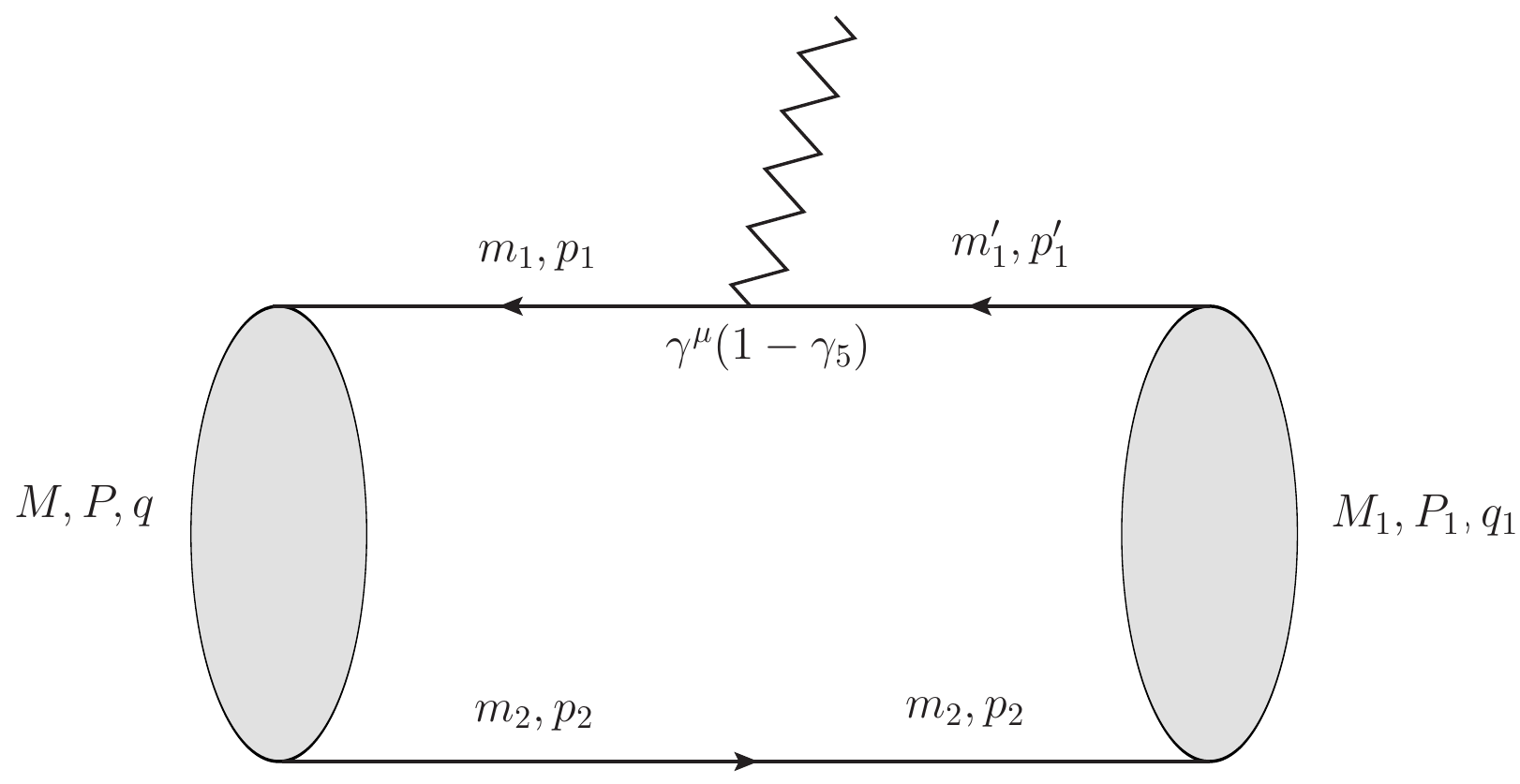}
\caption{Feynman Diagram of hadronic matrix element $M(P)\rightarrow M_1(P_1)$.}\label{fig:hm}
\end{figure}

This part will show the details how to calculate $\langle D^{*-}(P_1)\vert\bar{q}_2\gamma_\mu(1-\gamma_5)q_1\vert B^0(P)\rangle$ . The method is based on the instantaneous relativistic Bethe-Salpeter (BS) equation (Salpeter equation). We use $J^{P}$ to represent the meson, so $B^0$ is the $0^-$ meson and $D^{*-}$ is $1^-$ meson. The Feynman diagram of hadronic matrix element $\langle M_1\vert\bar{q}_2\gamma_\mu(1-\gamma_5)q_1\vert M\rangle$ is shown in Fig. \ref{fig:hm},
where $M$ is $0^-$ meson, $M_1$ is $1^-$ meson.

To get the numerical value of form factors $f$, $a_+$, $a_-$ and $g^\prime$ in Eq. (\ref{eq:hadronic}) we choose the Mandelstam formalism \cite{Mandelstam:1955sd}, where the transition amplitude between two mesons is described as a overlapping integral over the Bethe-Salpeter wave functions of initial and final mesons \cite{Chang:2006tc}. Using this method but with further instantaneous approximation, in center of mass system of initial meson, in leading order, we write the hadronic matrix element as:
\begin{equation}
\langle M_1(P_1)|{\bar{q}_1}\gamma^\mu(1-\gamma_5)q_2|M(P)\rangle=\int\frac{\mathrm{d}\vec{q}}{(2\pi)^3}\mathrm{Tr}
\left[\bar{\varphi}_{P_1}^{++}(\vec{q}_1)\gamma_\mu(1-\gamma_5)\varphi_P^{++}(\vec{q})\frac{\not\!P}{M}\right],\label{eq:amjuti}
\end{equation}
where $P$ and $P_1$ are the momenta of initial and final mesons; $M$ in denominator is the mass of initial meson; $q$ is relative momentum between quark and antiquark inside the initial meson, the relationship between them is
$p_1=\alpha_1 P+q,\alpha_1=\frac{m_1}{m_1+m_2}, p_2=\alpha_2 P-q,\alpha_2=\frac{m_2}{m_1+m_2}$; $\vec{q}_1=\vec{q}+\frac{m_{12}}{m_{11}+m_{12}}\vec{P}_1$ is the relative momentum inside the final meson $M_1$, $m_{11}$ ($m_{12}$) is mass of antiquark (quark) in final meson $M_1$, $\vec{P}_1$ is three dimensional momentum of meson $M_1$; $\varphi^{++}$ is the positive wave function for a meson in the BS method; for the final state, we have use the symbol $\bar{\varphi}_{P_1}^{++}=\gamma_0(\varphi_{P_1}^{++})^{+}\gamma_0$.

In the BS method, the positive wave function $\varphi^{++}$ for a vector can be written as \cite{Wang:2005qx}
\begin{eqnarray}\label{eq:1--}
\varphi^{++}_{1^-}&=&\frac{1}{2}\left[A\slashed\epsilon^{\lambda}_{1}+B\slashed\epsilon^{\lambda}_{1}\slashed P_1+C\left(\slashed q_{1P_\perp}\epsilon^{\lambda}_{1}-q_{1P_\perp}\cdot\epsilon^{\lambda}_{1}\right)+D\left(\right.\slashed P_1\slashed\epsilon^{\lambda}_{1}\slashed q_{1P_\perp}\right.\\\nonumber
&&\left.\left.-\slashed P_1q_{1P_\perp}\cdot\epsilon^{\lambda}_{1}\right)+q_{1P_\perp}\cdot\epsilon^{\lambda}_{1}(E+F\slashed P_1+G\slashed q_{1P_\perp}+H\slashed P_1\slashed q_{1P_\perp})\right],
\end{eqnarray}
where $\epsilon^{\lambda}_{1}$ is the polarization vector of the vector meson, $q_{1P_\perp}^{\mu}=q^{\mu}_1-q^{\mu}_{1P_\parallel}$, $q^{\mu}_{1P_\parallel}=\frac{P\cdot q_1}{M^{2}}P^{\mu}$ and
\begin{eqnarray}\label{eq:ABCDE}
A & = & M_1\left[f_5(\vec{q}_1)-f_6(\vec{q}_1)\frac{\omega_{11}+\omega_{12}}{m_{11}+m_{12}}\right]\nonumber,\\
B & = & \left[f_6(\vec{q}_1)-f_5(\vec{q}_1)
\frac{m_{11}+m_{12}}{\omega_{11}+\omega_{12}}\right]\nonumber,\\
C & = &
\frac{M_1(\omega_{12}-\omega_{11})}{m_{12}\omega_{11}+m_{11}\omega_{12}}\left[f_5(\vec{q}_1)
-f_6(\vec{q}_1)\frac{\omega_{11}+\omega_{12}}{m_{11}+m_{12}}\right]\nonumber,\\
D & = &
\frac{\omega_{11}+\omega_{12}}{\omega_{11}\omega_{12}+m_{11}
m_{12}+\vec{q}_1^2}\left[f_5(\vec{q}_1)
-f_6(\vec{q}_1)\frac{\omega_{11}+\omega_{12}}{m_{11}+m_{12}}\right]\nonumber,\\
E & = &
\frac{m_{11}+m_{12}}{M_1(\omega_{11}\omega_{12}+m_{11}
m_{12}-\vec{q}_1^2)}
\left\{M^{2}_1\left[f_5(\vec{q}_1)-f_6(\vec{q}_1)
\frac{m_{11}+m_{12}}{\omega_{11}+\omega_{12}}\right]\right.\nonumber\\
& &
-\left.\vec{q_1}^2\left[f_3(\vec{q_1})+f_4(\vec{q_1})
\frac{m_{11}+m_{12}}{\omega_{11}+\omega_{12}}\right]\right\},\nonumber\\
F & = &
\frac{\omega_{11}-\omega_{12}}{M^{2}_1(\omega_{11}\omega_{12}+m_{11}
m_{12}-\vec{q}_1^2)}
\left\{M^{2}_1\left[f_5(\vec{q}_1)-f_6(\vec{q}_1)
\frac{m_{11}+m_{12}}{\omega_{11}+\omega_{12}}\right]\right.\nonumber\\
& &
-\left.\vec{q}^2_1\left[f_3(\vec{q}_1)+f_4(\vec{q}_1)
\frac{m_{11}+m_{12}}{\omega_{11}+\omega_{12}}\right]\right\},\nonumber\\
G & = &
\left\{\frac{1}{M_1}\left[f_3(\vec{q}_1)+f_4(\vec{q}_1)
\frac{m_{11}+m_{12}}{\omega_{11}+\omega_{12}}\right]
-\frac{2f_6(\vec{q}_1)M_1}{m_{12}\omega_{11}+m_{11}\omega_{12}}\right\},\nonumber\\
H & = & \frac{1}{M^{2}_1}\left\{\left[f_3(\vec{q}_1)
\frac{\omega_{12}+\omega_{12}}{m_{11}+m_{12}}+f_4(\vec{q}_1)\right]
-2f_5(\vec{q}_1)\right.\nonumber\\
& &
\times\left.\frac{M^{2}_1(\omega_{11}+\omega_{12})}{(m_{11}+m_{12})(\omega_{11}\omega_{12}+m_{11}
m_{12}+\vec{q}_1^2)}\right\},
\end{eqnarray}
In Eq.\eqref{eq:ABCDE}, $m_{11}$ and $m_{12}$ are the masses of quark and antiquark inside the $1^-$ meson, and we list their values in Table~\ref{quark};
\begin{table}[ph]
\caption{ Mass of quark in unit of GeV. }
\begin{center}
{\begin{tabular}{|c|c|c|c|c|c|} \hline quark & $b$ & $c$ & $s$ & $d$ & $u$\\
\hline mass&$4.96$&$1.62$&$0.5$&$0.311$&$0.305$\\
\hline
\end{tabular} \label{quark}}
\end{center}
\end{table}
$\omega_{1i}$ is defined as
$\omega_{1i}=\sqrt{m_{1i}^{2}+\vec q^{2}_{1}}$, $i=1,2$; $f_3(\vec q_1), f_4(\vec q_1), f_5(\vec q_1), f_6(\vec q_1)$ are the radial wave functions of the meson.
Numerical values of wave functions $f_3(\vec q_1), f_4(\vec q_1), f_5(\vec q_1), f_6(\vec q_1)$ can be obtained by solving the coupled Salpeter equations \cite{Wang:2005qx}:

The positive wave function $\varphi^{++}$ for a pseudoscalar can be written as \cite{Wang:2009as}.
\begin{equation}\label{eq:0--}
\varphi^{++}_{0^-}=L\left(N+\frac{\slashed P}{M}+{\slashed q}_{P_\perp}Y+\frac{{\slashed q_{P_\perp}}{\slashed P}}{M}Z\right)\gamma_5,
\end{equation}
where the definition of $q_{\perp}$ is $q_{P_\perp}^{\mu}=q^{\mu}-q^{\mu}_{P_\parallel}$, $q^{\mu}_{P_\parallel}=\frac{P\cdot q}{M^{2}}P^{\mu}$, and
\begin{eqnarray}\label{eq:ABCD}
L&=&\frac{M}{2}\left[f_1(\vec{q})+f_2(\vec{q})\frac{m_1+m_2}{\omega_1+\omega_2}\right],\nonumber\\
N&=&\frac{\omega_1+\omega_2}{m_1+m_2},\nonumber\\
Y&=&-\frac{m_1-m_2}{m_1\omega_2+m_2\omega_1},\\
Z&=&\frac{\omega_1+\omega_2}{m_1\omega_2+m_2\omega_1}.\nonumber
\end{eqnarray}

Numerical values of wave functions $f_i({\vec{q})~(i=1,2)}$ can be obtained by solving the coupled Salpeter equations \cite{Wang:2009as}. Where we have chosen the Cornell potential, which is a linear potential plus a single gluon exchange reduced vector potential, and in momentum space the expression is:
\begin{eqnarray}
V_s(\vec{q})&=&-\left(\frac{\lambda}{\alpha}+V_0\right)\delta^3(\vec{q})+\frac{\lambda}{\pi^2}\frac{1}{(\vec{q}^2+\alpha^2)^2},\nonumber\\
V_v(\vec{q})&=&-\frac{2}{3\pi^2}\frac{\alpha_s(\vec{q})}{(\vec{q}^2+\alpha^2)},\nonumber\\
\alpha_s(\vec{q})&=&\frac{12\pi}{27}\frac{1}{\ln(a+\frac{\vec{q}^2}{\Lambda_{QCD}^2})},
\end{eqnarray}
where $a=e=2.71828$; $\lambda=0.21~\mathrm{GeV}^2$ is the string constant; $\alpha=0.06~\mathrm{GeV}$ is a parameter to avoid the infrared divergence; the QCD scale $\Lambda_{QCD}=0.27~\mathrm{GeV}$ characterize the running strong coupling constant $\alpha_s$; the constant $V_0$ is a parameter by hand in potential model to match the experimental dada, whose values for $B^0$ and $D^{*-}$ are $0.1$ and $0.11$.

With Eq. \eqref{eq:1--} and Eq. \eqref{eq:0--}, we take the integral on the right side of Eq. \eqref{eq:amjuti}, then the form factor $f$, $a_+$, $a_-$ and $g^\prime$ can be expressed as:
\begin{eqnarray}
f & = & T_1+T_{43},\nonumber\\
a_+ & = & \frac{1}{2}(T_2+T_2^\prime+T_{41}+T_{41}^\prime+T_3+T_3^\prime+T_{42}+T_{42}^\prime),\nonumber\\
a_- & = & \frac{1}{2}(T_2+T_2^\prime+T_{41}+T_{41}^\prime-T_3-T_3^\prime-T_{42}-T_{42}^\prime),\nonumber\\
g & = & \frac{1}{2}(M_1-M_2+M_3+M_4-M_5+M_6+M_7+M_8-M_9-M_{10}\nonumber\\
& & -M_{11}+M_{12}-M_{13}-V_1+V_2+V_3+V_4),
\end{eqnarray}
$t$, $T$, $M$ and $V$ can be written as:
\begin{eqnarray}
t_1 & = & A-BNE_1+BZ\vec P_1\cdot\vec
q-CZ(\vec{q}^2+\alpha_{12}\vec P_1\cdot\vec
q)\nonumber\\
& & +D(\alpha_{12}\vec{P_1}^2+\vec P_1\cdot\vec q)-D
YE_1(\vec{q}^2+\alpha_{12}\vec P_1\cdot\vec
q),\nonumber\\
t_2 & = & -\frac{AY}{M}+\frac{DY}{M}(\alpha_{12}
M^{2}_1-\vec P_1\cdot\vec q)-\frac{EN}{M}+\frac{FY}{M}\vec P_1\cdot\vec q\nonumber\\
& & +\frac{GY}{M}(\vec{q}^2+\alpha_{12}\vec P_1\cdot\vec
q)-\frac{HN}{M}(\alpha_{12} M^{2}_1-\vec P_1\cdot\vec
q)\nonumber\\
& & +\frac{CZ}{M}\alpha_{12}
E_1-\frac{G}{M}\alpha_{12} E_1+\frac{H
Z}{M}\alpha_{12} E_1\vec P_1\cdot\vec q,\nonumber\\
t_2^\prime & = & \frac{D Y\alpha_{12}
E_1}{M^2}\vec P_1\cdot\vec q-\frac{F Y\alpha_{12}
E_1}{M^2}\vec P_1\cdot\vec q-\frac{GY\alpha_{12}
E_1}{M^2}(\vec{q}^2+\alpha_{12}\vec P_1\cdot\vec
q)\nonumber\\
& & +\frac{HN\alpha_{12}E_1}{M^2}(\alpha_{12} M^{2}_1-\vec P_1\cdot\vec q)-\frac{H Z\alpha_{12} E_1^2}{M^2}\vec P_1\cdot\vec q\nonumber\\
& & +\frac{E N\alpha_{12}E_1}{M^2}+\frac{C N\alpha_{12}E_1}{M^2}+\frac{G\alpha_{12} E_1^2}{M^2},\nonumber\\
t_3 & = & BZ-D Y\alpha_{12} E_1+F+HZ\vec{q}^2-C Z\alpha_{12}+G\alpha_{12}
+HN\alpha_{12} E_1,\nonumber\\
t_3^\prime & = & \frac{BN}{M}+\frac{DY}{M}\vec{q}^2-\frac{F}{M}\alpha_{12} E_1 -\frac{CN}{M}\alpha_{12}-\frac{HZ}{M}\alpha_{12}E_1\vec{q}^2\nonumber\\
& & -\frac{G}{M}\alpha_{12}E_1
-\frac{H N}{M}\alpha_{12}E_1^2,\nonumber\\
t_4 & = & -DYE_1-EZ+FYE_1-CZ
+G+HZ\alpha_{12}\vec{P}_1^2+HNE_1,\nonumber\\
t_4^\prime & = & \frac{AY}{M}+\frac{DY}{M}\alpha_{12}\vec{P}_1^2+\frac{EZ}{M}\alpha_{12}E_1-\frac{CN}{M}-\frac{FY}{M}\alpha_{12}E_1^2\nonumber\\
& & -\frac{HZ}{M}\alpha_{12}E_1\vec{P}_1^2-\frac{G}{M}\alpha_{12} E_1-\frac{HN}{M}\alpha_{12} E_1^2,
\end{eqnarray}
\begin{eqnarray}
T_1 & = & 4A\int\frac{d{\vec q}}{(2\pi)^3}t_1,\nonumber\\
T_2 & = & -4A\frac{E_1}{M|\vec P_1|}\int\frac{d{\vec
q}}{(2\pi)^3}|\vec
q|\cos\theta t_2,\nonumber\\
T_2^\prime & = & 4A\int\frac{d{\vec q}}{(2\pi)^3}t_2^\prime,\nonumber\\
T_3 & = & -4A\frac{E_1}{M|\vec P_1|}\int\frac{d{\vec
q}}{(2\pi)^3}|\vec
q|\cos\theta t_3,\nonumber\\
T_3^\prime & = & 4A\int\frac{d{\vec q}}{(2\pi)^3}t_3^\prime,\nonumber\\
T_{41} & = & 2A\frac{1}{M^2|\vec P_1|^2}\int\frac{d{\vec
q}}{(2\pi)^3}|\vec q|^2\left[(M^{2}_1+2E_1^2){\cos}^2\theta-M^{2}_1\right]t_4,\nonumber\\
T_{41}^\prime & = & -4A\frac{E_1}{M|\vec{P}_1|}\int\frac{d{\vec
q}}{(2\pi)^3}|\vec
q|\cos\theta t_4^\prime,\nonumber\\
T_{42} & = & -2A\frac{E_1}{M|\vec P_1|^2}\int\frac{d{\vec
q}}{(2\pi)^3}|\vec q|^2(3{\cos}^2\theta-1)t_4,\nonumber\\
T_{42}^\prime & = & 4A\frac{1}{|\vec P_1|}\int\frac{d{\vec
q}}{(2\pi)^3}|\vec
q|\cos\theta t_4^\prime,\nonumber\\
T_{43} & = & 2A\int\frac{d{\vec q}}{(2\pi)^3}|\vec
q|^2({\cos}^2\theta-1)t_4,
\end{eqnarray}
\begin{eqnarray}
M_1 & = & -4L\frac{1}{M|\vec P_1|}\int\frac{d{\vec q}}{(2\pi)^3}|\vec
q|\cos\theta AY,\nonumber\\
M_2 & = & -4L\frac{E_1}{M|\vec P_1|}\int\frac{d{\vec
q}}{(2\pi)^3}|\vec
q|\cos\theta B Z,\nonumber\\
M_3 & = & 4L\frac{1}{M|\vec P_1|}\int\frac{d{\vec q}}{(2\pi)^3}|\vec
q|\cos\theta C N,\nonumber\\
M_4 & = & -4L\frac{\alpha_{12} E_1}{M|\vec P_1|}\int\frac{d{\vec
q}}{(2\pi)^3}|\vec
q|\cos\theta CZ,\nonumber\\
M_5 & = & -4L\frac{\alpha_{12} E_1}{M|\vec P_1|}\int\frac{d{\vec
q}}{(2\pi)^3}|\vec
q|\cos\theta CZ,\nonumber\\
M_6 & = & 4L\frac{E_1}{M|\vec P_1|}\int\frac{d{\vec q}}{(2\pi)^3}|\vec
q|\cos\theta D\nonumber\\
M_7 & = & 4L\frac{\alpha_{12} M^{2}_1}{M|\vec
P_1|}\int\frac{d{\vec q}}{(2\pi)^3}|\vec
q|\cos\theta DY,\nonumber\\
M_8 & = & -4L\frac{\alpha_{12} E_1^2}{M|\vec P_1|}\int\frac{d{\vec
q}}{(2\pi)^3}|\vec
q|\cos\theta DY,\nonumber\\
M_9 & = & 2L\frac{1}{M}\int\frac{d{\vec
q}}{(2\pi)^3}|\vec q|^2({\cos}^2\theta-1)DY,\nonumber\\
M_{10} & = & -2L\frac{1}{M}\int\frac{d{\vec
q}}{(2\pi)^3}|\vec q|^2({\cos}^2\theta-1)FY,\nonumber\\
M_{11} & = & -2L\frac{\alpha_{12}}{M}\int\frac{d{\vec
q}}{(2\pi)^3}|\vec q|^2({\cos}^2\theta-1)GY,\nonumber\\
M_{12} & = & 2L\frac{1}{M}\int\frac{d{\vec
q}}{(2\pi)^3}|\vec q|^2({\cos}^2\theta-1)HN,\nonumber\\
M_{13} & = & -2L\frac{\alpha_{12} E_1}{M}\int\frac{d{\vec
q}}{(2\pi)^3}|\vec q|^2({\cos}^2\theta-1)HZ,
\end{eqnarray}
\begin{eqnarray}
V_1 & = & 4L\frac{1}{M}\int\frac{d{\vec q}}{(2\pi)^3}BN,\nonumber\\
V_2 & = & 4L\frac{\alpha_{12}}{M}\int\frac{d{\vec q}}{(2\pi)^3}CN,\nonumber\\
V_3 & = & 4L\frac{\alpha_{12}E_1}{M}\int\frac{d{\vec q}}{(2\pi)^3}D,\nonumber\\
V_4 & = & -4L\frac{1}{M}\int\frac{d{\vec q}}{(2\pi)^3}|\vec
q|^2DY,
\end{eqnarray}
where $\alpha_{12}=\frac{m_{12}}{m_{11}+m_{12}}$, $\cos{\theta}=\frac{\vec P_1\cdot\vec q}{|\vec{P_1}||\vec{q}|}$.
\bibliography{bibliography}
\end{document}